\documentclass[twocolumn,showpacs,preprintnumbers,amsmath,amssymb]{revtex4}
\usepackage{graphicx}
\usepackage{dcolumn}
\usepackage{bm}

\begin{document}
\preprint{APS}
\title{Numerical investigation of the quantum fluctuations of optical fields transmitted through an atomic medium}
\author{A. Lezama$^{1}$, P. Valente$^{2}$, H. Failache$^{1}$, M. Martinelli$^{2}$ and P.
Nussenzveig$^{2}$}
\affiliation{$^{1}$ Instituto de F\'{i}sica, Facultad de
Ingenier\'{i}a. Casilla de correo 30, 11000, Montevideo, Uruguay}
\affiliation{$^{2}$ Instituto de F\'{i}sica, Universidade de
S\~{a}o Paulo, Caixa Postal 66318, 05315-970, S\~{a}o Paulo, SP,
Brazil }
\date{\today}

\begin{abstract}
We have  numerically solved the Heisenberg-Langevin equations
describing the propagation of quantized fields through an
optically thick sample of atoms. Two orthogonal polarization
components are considered for the field and the complete Zeeman
sublevel structure of the atomic transition is taken into account.
Quantum fluctuations of atomic operators are included through
appropriate Langevin forces. We have considered an incident field
in a linearly polarized coherent state (driving field) and vacuum
in the perpendicular polarization and calculated the noise spectra
of the amplitude and phase quadratures of the output field for two
orthogonal polarizations. We analyze different configurations
depending on the total angular momentum of the ground and excited
atomic states. We examine the generation of squeezing for the
driving field polarization component and vacuum squeezing of the
orthogonal polarization. Entanglement of orthogonally polarized modes
is predicted. Noise spectral features specific of (Zeeman) multi-level
configurations are identified.
\end{abstract} 

\maketitle
\section{\label{introduccion}Introduction}
The preparation of optical fields in states with purely
quantum-mechanical properties is the key ingredient of quantum
optics and the essential requirement for their use in quantum
information processing. Light fields presenting squeezing  are
well known examples of non-classical states for which
numerous applications have been suggested and demonstrated.
Entanglement of a two-mode field is another 
important example of a purely quantum mechanical resource which
lies at the basis of a large number of quantum information
procedures such as EPR pair production, teleportation, and quantum
cryptography \cite{TERHAL03}. For two degenerate field modes,
considered as continuous variable systems, 
squeezing and entanglement are related concepts. Squeezing in one
mode leads to entanglement between two modes obtained by a linear
optical transformation, as implemented by a beam
splitter~\cite{BOHR35,KIM02}.

Ever since the first proposals for light squeezing, atomic systems
have attracted considerable attention owing to the large
nonlinearities associated to resonant transitions. Indeed, the
first successful demonstration of squeezing used non-degenerate
four-wave mixing in sodium atoms \cite{SLUSHER85} contained in an
optical cavity. Several subsequent experiments
\cite{RAIZEN87,HOPE92} also used atomic samples contained in
optical cavities. In these experiments, the highly nonlinear
interaction between the atoms and the cavity-mode plays an
essential role in the generation of squeezing.

Following the early work, most present day experiments on the
generation of non-classical fields with atomic samples involve the
use of optical cavities and require rather complicated
experimental setups \cite{LAMBRECHT96,JOSSE03,JOSSE04}. However, in
view of applications, the use of single path schemes for the
generation of non-classical light fields could be of considerable
practical interest. In this paper, we are concerned with the
modification of quantum fluctuations of a single monochromatic
light beam interacting with an atomic medium on a single path.
Such possibility  is already implicit in the pioneering work by Walls
and Zoller \cite{WALLS81} and Mandel \cite{MANDEL82} predicting
reduced quantum fluctuations in the light emitted by a resonantly
driven two-level atom. The spectral distribution of the quadrature
fluctuations of light emitted by a driven two-level atom was
first calculated in \cite{COLLETT84}. The
generalization of this study to an extended atomic sample was
carried out by Heidmann {\em et al.} \cite{HEIDMANN85}, who considered the
fluctuations in the field emitted in the forward direction by a
thin layer of atoms at rest, driven by a normally incident laser
beam. At low laser intensities, squeezing is predicted for the low
frequency components of the in-phase quadrature. For saturating
intensities, squeezing occurs for noise frequencies around the
generalized Rabi frequency
$\Omega=(\Delta^{2}+\Omega_0^{2})^{1/2}$, where $\Delta$ is the
atom-laser detuning, and $\Omega_0$ the incident field resonant
Rabi frequency. The calculation of the fluctuation spectra of
light traversing a thick two-level atom medium was presented by Ho
{\em et al.} \cite{HO87}. Single path squeezing  was observed using
sodium \cite{HO91} and ytterbium \cite{LU98}. Quite recently,
intensity-intensity quantum correlations (sub-shot noise intensity-difference
noise) were observed in non-degenerate forward four-wave mixing
in rubidium vapor \cite{MCCORMICK07}.

The initial work on squeezing through light-atom interaction
considered ideal two-level transitions and a single-mode optical
field. A more realistic approach requires the consideration of
multi-mode light fields (including different polarizations) and
multi-level atoms. Three-level atoms interacting with two fields
($\Lambda$ system) have been analyzed. Such systems bear a
considerable interest owing to the possibility of large
nonlinearities in association with electromagnetically induced
transparency (EIT). Phase-noise squeezing was predicted for a
$\Lambda$ system in which one of the fields was classical, for a
cavity contained atomic system \cite{FLEISCHHAUER92} and for
single path propagation through an ensemble of motionless atoms
\cite{RATHE96}.

An interesting issue of the multi-mode field interaction with an
atomic sample is the possibility to achieve polarization
squeezing, which is signalled by squeezing of the vacuum field with orthogonal
polarization relative to the incident field. Polarization
squeezing with cold atoms inside an optical cavity was
demonstrated by Josse and co-workers \cite{JOSSE03,JOSSE03BIS}.
Polarization squeezing is intimately related to the observation of
continuous variable entanglement between two field modes
\cite{JOSSE04,DANTAN06,PIELAWA07,DUAN00}.

Recently, a renewal of attention on the fluctuations of light
transmitted through an atomic sample was motivated by the
prediction that vacuum field squeezing could be
achieved on a single path as a consequence of polarization
self-rotation (PSR) \cite{MATSKO02}. An experimental observation
of squeezing via PSR was reported \cite{RIES03}, using a room
temperature rubidium cell. Such result could not be reproduced by
other groups, in spite of the methodical exploration of the
relevant experimental parameter space \cite{HSU06}. It is argued
by Hsu and co-workers \cite{HSU06} that excess noise, preventing
the observation of squeezing via PSR, originates from the quantum
atomic fluctuations not explicitly included in the original
theoretical proposal \cite{MATSKO02}. Such argument is supported
by a simplified four-level system model \cite{JOSSE03BIS}, for
which the noise arising from atomic quantum fluctuations dominates
over the semiclassical squeezing terms, under the conditions
corresponding to the experiments. However, PSR squeezing is not
excluded for cold atom samples in the regime of large intensities
and optical detunings.

The noise properties of light fields interacting with atom samples
on a single path have been experimentally investigated under
conditions of EIT. Large correlations and anti-correlations were
observed between the two fields at the Raman resonance condition
between the two ground state hyperfine levels of Rb
\cite{GARRIDO03,CRUZ07}. The change in sign of the correlation is
determined by the light intensity. Noise spectra and correlations
between two polarization components participating in Hanle/EIT
resonance involving Zeeman sublevels of the Rb ground state were
observed \cite{MARTINELLI04}. These studies were carried out using
diode lasers, known to possess large excess phase noise. A
qualitative agreement between these experiments and theory could
be reached by considering a classical field with random phase
diffusion.

All theoretical models considered so far in connection with the
analysis of light fluctuations interacting with an atomic sample
are based on several simplifications. Single-mode approaches
ignore the light polarization orthogonal to the incident field,
through which vacuum fluctuations enter the atomic system. Such
fluctuations interact with the incident field,
provided the sample possesses or acquires some anisotropy 
through interaction with light. 
 Multi-mode models also rely on
simplifying assumptions. Sometimes one of the incident fields is
taken as classical \cite{FLEISCHHAUER92}. The transverse spatial
structure of the field is generally not considered
\cite{LAMBRECHT96BIS}. In most 
cases, quantum fluctuations of the
atomic operators are ignored.
 Even when a full
quantum treatment was used, the atomic level structure considered
was assumed to be an ideal three or four-level system. In most
studies, the effect of propagation through an optically thick
sample is not examined. The effect of the atomic velocity
distribution and that of nearby atomic levels within the hyperfine
structure is usually neglected.

It is the purpose of this work to investigate the quantum noise in
fields transmitted through an homogeneous atomic sample when the
full Zeeman degeneracy of the ground and excited atomic levels is
taken into account. We consider a light field incident on the
atomic sample with a well defined polarization and take into
account the incoming vacuum field fluctuations with orthogonal
polarization. We calculate the field fluctuation noise spectrum
after propagation through the medium for arbitrary polarization
and field quadrature angle, fully taking into account the influence
of the quantum fluctuations of the atomic medium. Depending on the
choice of the ground and excited level angular momenta ($F_{g}$
and $F_{e}$ respectively) and the field polarization, several
configurations can be analyzed. A two-level system is obtained for
a $F_{g}=0 \rightarrow F_{e}=1$ transition. An open $\Lambda$
system is obtained for a $F_{g}=1 \rightarrow F_{e}=0$ transition,
if the incident field is linearly polarized and the two circular polarization
components are considered. The four level system studied in
\cite{JOSSE03BIS} corresponds to a $F_{g}=1/2 \rightarrow
F_{e}=1/2$ transition, with linear field polarization. In addition
to those schemes previously explored, our calculation allows us to
address other configurations, such as the $F_{g}\neq 0 \rightarrow
F_{e}=F_{g}+1$, for which no full quantum treatment was previously
reported in spite of the prediction of PSR squeezing in this
system \cite{MATSKO02}. Our calculation considers a 1-dimensional
propagation (along the $z$ axis) through a spatially homogeneous
atomic sample considered, as a continuous medium
\cite{FLEISCHHAUER95}. We begin by assuming that the atoms are at
rest. Next, we briefly address the effect of a thermal atomic
velocity distribution. Propagation effects on quantum
fluctuations for a thick medium, of length $L$, are included in the
calculation, although the depletion of the incident field mean
value is ignored for simplicity. Thus, our results can be
directly applied to situations in which absorption of the
carrier frequency component of the field can be neglected
(saturating field intensities or large atom--field detunings) . \\

Our calculation confirms the prediction of quadrature squeezing
under several conditions and indicates that squeezing by PSR can
take place for $F_{g} =1/2 \rightarrow F_{e}=1/2$ transitions,
and for $F_{g}> 0 \rightarrow F_{e}=F_{g}+1$ transitions as well.
The rather complex structure of the noise spectrum for thick optical
media is illustrated and new features in the noise spectrum specific
of the multi-level Zeeman structure are presented.\\

The paper is organized as follows. In the next section, the main
lines of our theoretical model and noise spectrum calculation are
presented. In section \ref{resultados}, we present and discuss the
noise spectra corresponding to different level schemes. Concluding
remarks are made in section \ref{conclusiones}.\\

\section{\label{modelo}Model}

\begin{figure}
\includegraphics[width=8.6cm]{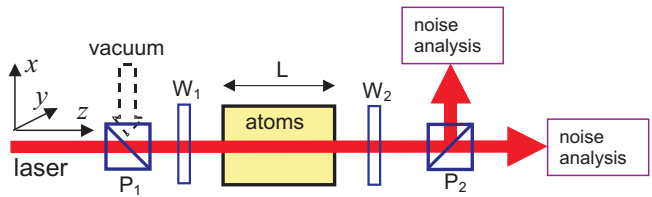}
\caption{\label{setup} (Color online) Scheme of the physical
situation theoretically addressed in this paper ($P_{1}$, $P_{2}$:
polarizers, $W_{1}$, $W_{2}$: waveplates).  The driving field
polarization is imposed by $P_{1}$ and $W_{1}$, while $W_{2}$ and
$P_{2}$ achieve the decomposition of the transmitted field into
orthogonal polarization components. The noise analysis is made by
homodyne detection. In the calculations, we have chosen linear $x$
polarization for the driving field and analyzed fluctuations for
polarizations $x$ and $y$.}
\end{figure}

The problem that we address in this paper is schematically
presented in Fig. \ref{setup}. A laser beam traverses an optically
thick atomic sample. The polarization of the incident driving
field is defined by appropriate polarization optics. After the
sample, the field is decomposed into two chosen 
orthogonal polarizations. The quadrature noise on both polarization
components is analyzed by homodyne detection \cite{BACHORBOOK98}.

Our theoretical approach follows closely the method  used by
Dantan and co-workers \cite{DANTAN05} for the study of the
propagation of light fluctuations through an ensemble of three-level
atoms in a $\Lambda $ configuration. We consider two
levels with Zeeman sublevels: a ground state $g$, of total angular momentum
$F_g$ and zero energy, and an excited state $e$, of angular momentum
$F_e$ and energy $\hbar \omega _0$. The total radiative relaxation
coefficient of level $e$ is $\Gamma $. We assume that the atoms in
the excited state can radiatively decay into the ground state $g\
$at a rate $b\Gamma $, where $b$ is a branching ratio coefficient
that depends on the specific atomic transition $ (0\leq b\leq 1)
$.  For a closed (cycling) transition, $b=1$. When the transition
is open ($b<1$), excited atoms can decay back into level $g$ or
into levels external to the two-level system. The atoms are in the
presence of a magnetic field $\mathcal{B}$ directed along the
light propagation axis $z$. In order to simulate the effect of a
finite interaction time of the atoms with the light, we introduce
an overall phenomenological decay constant $\gamma $, $(\gamma \ll
\Gamma)$ which is compensated, in the steady-state, by the arrival of
fresh atoms in the ground state. We initially analyze an
homogeneous ensemble of atoms at rest, leaving the effect of the atomic
velocity distribution for subsequent consideration.

Incident upon the atoms is a light field described in the
Heisenberg picture by the operator:

\begin{eqnarray}
\vec{E} (z,t) &=&\xi  \left( a_1e^{i(kz-\omega _Lt)}%
\hat{e}_1^{*}+a_2e^{i(kz-\omega _Lt)}\hat{e}_2^{*}\right.
\label{campo} \\
&&\left. +a_1^{\dagger }e^{-i (kz-\omega _Lt)  }\hat{e}%
_1+a_2^{\dagger }e^{-i (kz-\omega _Lt)  }\hat{e}_2 \right)\;,
\nonumber
\end{eqnarray}
where $\hat{e}_1$ and $\hat{e}_2
$ are two orthogonal (complex) polarization unit vectors and $a_1$, $a_2$, $%
a_1^{\dagger }$, $a_2^{\dagger }$ are the slowly varying field
operators obeying the commutation rules $\left[ a_\kappa (z,t)
,a_\lambda (z^{\prime },t^{\prime })  \right] =0$ and $\left[
a_\kappa (z,t) ,a_\lambda^{\dagger } (z^{\prime },t^{\prime })
\right] =\delta _{\kappa \lambda}\frac Lc\delta (t-t^{\prime
}-\frac{z-z^{\prime }}c)$. The quantization length $L$
is chosen as the atomic medium length and $c$ is the speed of light in
vacuum. $\xi =\sqrt{\frac{\hbar \omega_{L}}{2\epsilon_{0} SL}}$ is
the single photon field amplitude (where $S$ is the mode cross-section).

The atomic operators in the Heisenberg representation for an atom
$j$ at position $z_j$ are $\rho _{\mu \nu }^j (t)  =\left| \nu
\right\rangle \left\langle \mu \right| _j$ where $\left| \mu
\right\rangle$ and $\left| \nu \right\rangle$ designate Zeeman
sub-states. We introduce the slowly varying atomic operators
$\sigma _{\mu \nu }^j (z_j,t) =U\rho _{\mu \nu }^j (t) U^{\dagger
}$where $U=P_e^je^{i (kz_j-\omega _Lt)  }+P_g^j$ is an unitary
transformation. $P_g^j$ and $P_e^j$ are the projectors on ground
and excited state manifolds respectively. Following
\cite{DANTAN05}, we define continuous local operators (at position
$z)$ by averaging over a slice of the atomic medium of length
$\Delta z$:

\begin{equation}
\sigma _{\mu \nu } (z,t)  =\lim_{\Delta z\rightarrow 0}\frac
L{N\Delta z}\sum_{z\leqslant z_j\leqslant z+\Delta z}\sigma _{\mu
\nu }^j (z_j,t)\;,  \label{sigma de z}
\end{equation}
where $L$ is the total length of the atomic medium and $N$ the
number of atoms.

The atomic Hamiltonian is:
\begin{equation}\label{hamiltoniano atomico}
H_A=\frac NL\int  (H_0+H_B)  dz\;,
\end{equation}
with $H_0=\hbar \omega _0P_e$ the isolated atom Hamiltonian and
$H_B=(\beta _gP_g+\beta _eP_e)F_z\mathcal{B}$ the Zeeman coupling
with the magnetic field $\mathcal{B}$. $P_e $ and $P_g$ are the
local projectors on the excited and ground manifolds, respectively,
$\beta _g$ and $\beta _e$ are the ground and excited state
gyromagnetic factors, and $F_z$ is the local total angular momentum
operator projection along the magnetic field axis $z$.

The atom field coupling $H_{Int}$, in the rotating wave
approximation, is:
\begin{equation}
H_{Int}=-\frac NL\hbar \eta\int \left[  (a_1^{\dagger }\hat{e}_1+a_2^{\dagger }%
\hat{e}_2)  .\vec{Q}_{ge}+h.c.\right] dz \;. \label{Interaccion atomo
campo}
\end{equation}

Here $\vec{Q}_{ge}=(\vec{Q}_{eg})^{\dagger }=P_g\vec{Q}P_e$ is a
dimensionless operator related to the atomic electric dipole
operator $\vec{D}$ through $\vec{D}=\left\langle g\Vert
\vec{D}\Vert e\right\rangle \vec{Q}$. The reduced matrix element
$\left\langle g\Vert\vec{D}\Vert e\right\rangle$ of the
dipole operator between the ground and excited state is taken to be real.
$\eta = \xi \left\langle g\Vert \vec{D}\Vert e\right\rangle
/\hbar$ is the \emph{reduced} atom--field coupling constant (half the single
photon Rabi frequency). The standard spherical components of the
operator $\vec{Q}_{eg}$ are $Q_{eg}^q$, with ($q=-1,0,1$). Their
matrix elements $Q_{\mu \nu,ge}^q\equiv \left\langle
\mu\right| Q_{ge}^q$ $\left| \nu\right\rangle $ (where $\mu$ and
$\nu$ refer to Zeeman substates belonging to the ground and
excited manifolds respectively) are the corresponding
Clebsch-Gordan coefficients.

The complete set of atomic operators can be organized into a
two-dimension operator array $\sigma \equiv \left\{ \sigma
_{ij}\right\} $ whose elements are the individual $\ \sigma _{ij}$
operators. After manipulation, one can formally write the
Heisenberg-Langevin equations (including relaxation terms) for the
atomic operators in the form:

\begin{eqnarray}
d\sigma /dt &=&-i \Delta \left[ P_e,\sigma \right]
-\frac i\hbar \left[ H_B,\sigma \right] \nonumber \\
&&+i\eta\left[ \left\{  (a_1\hat{e}_1^{*}+a_2\hat{e}_2^{*})  .%
\vec{Q}_{eg} + h.c. \right\} ,\sigma \right]  \nonumber \\
&& +b\Gamma  (2F_e+1)  \sum_qQ_{ge}^q\sigma Q_{eg}^q \nonumber \\
&&-\frac \Gamma 2\left\{ P_e,\sigma \right\} -\gamma  (\sigma
-\sigma ^0)  +f \;, \label{HL con operador sigma}
\end{eqnarray}
where $\Delta =\omega _0 -\omega _L$ is the optical detuning and
$\gamma \sigma_{0}$ is a pumping term describing the isotropic
arrival of fresh atoms in the lower ground state. $\sigma ^0\equiv
\left\{ \sigma _{ij}^0\right\} $ is chosen such that $\sigma_{ij}^0
= 0$ for $i\ne j$, $\sigma_{ii}^0=\frac{\mathbb{I}}{ (2F_g+1) }$ ($\mathbb{I}$
is the identity operator) if state $i$ belongs to the ground level, and
$\sigma _{ij}^0=0$ otherwise. $f\equiv\left\{f_{ij}\right\} $ represents
the set of Langevin force operators satisfying \cite{DANTAN05}: $%
\left\langle f_{ij} (z,t)  f_{kl}^{\dagger } (z^{\prime
},t^{\prime })  \right\rangle =\frac LN2D_{ij,kl}\delta
 (z^{\prime }-z^{\prime \prime })  \delta  (t-t^{\prime }) \;, $ where $D_{ij,kl}$ is the corresponding
diffusion coefficient.

The field evolution is governed by the Maxwell-Heisenberg
equations (in the slowly varying envelope approximation):

\begin{subequations}
\label{Maxwell}
\begin{eqnarray}
 (\frac \partial {\partial t}+c\frac \partial {\partial
z})  a_\lambda &=&iN\eta\hat{e}_\lambda.\sum_{_{_{k\in e,l\in
g}}}\vec{Q}_{ge,lk}\sigma _{kl} \\
 (\frac \partial {\partial t}+c\frac \partial {\partial
z})
a_\lambda^{\dagger } &=&-iN\eta\hat{e}_\lambda^{*}.\sum_{_{_{k\in e,l\in g}}}\sigma _{lk}%
\vec{Q}_{eg,kl}
\end{eqnarray}
\end{subequations}
with $\lambda=1,2$.

Eqs. \ref{HL con operador sigma} and \ref{Maxwell} form a set of
coupled differential equations for the atom--field interaction
along the atomic sample. We simplify the solution of these
equations by assuming that the process is stationary and that the
incident field mean value is undepleted as the atomic medium is
traversed $\left\langle a_\lambda (z)  \right\rangle =\left\langle
a_\lambda (0)  \right\rangle =\left\langle
a_\lambda\right\rangle $. We proceed by linearizing the field and atom operators: $%
\sigma _{ij} (z,t)  =\left\langle \sigma _{ij}\right\rangle
+\delta \sigma _{ij} (z,t)  $ and $a_\lambda (z,t)
=\left\langle a_\lambda\right\rangle +\delta a_\lambda (z,t)  $ with $%
\left\langle \delta \sigma _{ij}\right\rangle =\left\langle \delta
a_\lambda\right\rangle =0$.

The mean value of the atomic operators, given the incident field
mean values, can be obtained by taking the mean value of both
sides of Eq. \ref {HL con operador sigma}. This gives the usual
Bloch equations. A numerical solution of these equations was
presented in \cite{LEZAMA00}. We consider next the evolution of
the fluctuation operators up to first order. Fourier transforming
the first order contributions in Eqs. \ref{HL con operador sigma}
and  \ref{Maxwell} we get:

\begin{subequations}
\label{Maxwell-fluctuations}
\begin{eqnarray}
\frac \partial {\partial z}\delta a_j (z,\omega) &=&i\frac \omega %
c \delta a_j (z,\omega) \\
&&+iN\frac{\eta}{c}\hat{e}_j.\sum_{_{_{k\in e,l\in g}}}\vec{Q}_{ge,lk}\delta \sigma _{kl} (z,\omega) \nonumber \\
\frac \partial {\partial z}\delta a_j^{\dagger } (z,\omega) %
&=&i\frac \omega c\delta a_j^{\dagger } (z,\omega) \\
&&-iN\frac{\eta}{c} \hat{e}_j^{*}.\sum_{_{_{k\in e,l\in g}}}\delta
\sigma_{lk} (z,\omega)  \vec{Q}_{eg,kl} \nonumber
\end{eqnarray}
\end{subequations}

\begin{eqnarray}
f (\omega) &=&-i\omega \delta \sigma +i \Delta \left[ P_e,\delta
\sigma \right] +\frac i\hbar \left[ H_B,\delta \sigma \right]
\nonumber \\
&&-i\eta\left[ \left\{  (\left\langle a_1^{\dagger }\right\rangle \hat{e}%
_1+\left\langle a_2^{\dagger }\right\rangle \hat{e}_2)  .\vec{Q}%
_{ge}+h.c.\right\} ,\delta \sigma \right]   \nonumber \\
&&-i\eta\left[ \left\{  (\delta a_1^{\dagger }\hat{e}_1+\delta
a_2^{\dagger }\hat{e}_2)  .\vec{Q}_{ge}+h.c.\right\} ,\left\langle
\sigma \right\rangle \right]   \nonumber \\
&&-b\Gamma (2F_e+1)   (\sum_qQ_{ge}^q\delta \sigma Q_{eg}^q) \nonumber \\
&&+\frac \Gamma 2\left\{ P_e,\delta \sigma \right\} +\gamma \delta%
\sigma \;,\label{Heisenberg-fluctuations}
\end{eqnarray}
where $f (\omega)  \equiv \left\{ f_{ij} (z,\omega) \right\} $,
with $\left\langle f_{ij} (z,\omega) f_{kl}^{\dagger } (z^{\prime}
\omega^{\prime })  \right\rangle =\frac LN2D_{ij,kl}\delta
(z-z^{\prime })  \delta (\omega -\omega ^{\prime }) $.

Field fluctuations depend linearly on atomic fluctuations, which in
turn are driven by the Langevin force operators. To numerically
solve these equations, we adopt a Liouville-space approach
\cite{LEZAMA00,VERNAC02}, organizing all operators $\sigma _{ij}$
into a column vector $x$, with $n=4 (F_g+F_e+1)  ^2$ elements, and
the four field operators $a_1,a_1^{\dagger },a_2,a_2^{\dagger }$
into a four-element column vector $A$. Then, with some
manipulation, Eqs. \ref{Maxwell-fluctuations} and \ref
{Heisenberg-fluctuations} can be written in the form
(we drop the dependence on $z$ and $\omega $ for shortness):
\begin{equation}
\frac{\partial \delta A}{\partial z}=i\frac \omega c\Bbb{I}_4\delta A+\frac{%
N\eta}cW\delta x  \label{linearizada deltaa}
\end{equation}
\begin{equation}
- (i\omega \Bbb{I}_n+\mathcal{A})  \delta x=f+\eta V\delta A\;,
\label{linearizada deltax}
\end{equation}
where $\Bbb{I}_n$ is a $n\times n$ identity matrix, $W$ is a
$4\times n$
matrix dependent on the coefficients $Q_{ij}^q,$ $\mathcal{A}$ is a $%
n\times n$ matrix corresponding to the atomic evolution (including
relaxation terms), and $V$ is a $n\times 4$ matrix describing the
coupling of the field fluctuations to the atomic operator mean
value. By defining $M\equiv  -(i\omega \Bbb{I}_n+\mathcal{A})  $, we can
invert Eq. \ref{linearizada deltax} and from Eq. \ref{linearizada
deltaa} we get:
\begin{equation}
\frac{\partial \delta A}{\partial z}=B\delta A+\frac{N\eta}{c}Gf\;,
\label{propagg linearizada}
\end{equation}
with $G=WM^{-1}$ a $4\times n$ matrix and $B= (i\frac \omega c\Bbb{I}_4+%
\frac{N\eta^2}cGV)  $ a $4\times 4$ matrix. A formal solution of
Eq. \ref {propagg linearizada} for propagation over a length $z$
is given by:
\begin{eqnarray}
\delta A (z,\omega) &=&e^{Bz} \left[ \delta A (0,\omega) \right. \\
&&+\left. \frac{N\eta}c\int_0^ze^{-Bz^{\prime }}Gf (z^{\prime
},\omega) dz^{\prime }\right] \;.\nonumber \label{formal}
\end{eqnarray}

The power spectra of the fields' fluctuations after propagation
through an atomic medium of thickness $z$ can be obtained from the
matrix $S( z,\omega ) $ related to the field operators' spectral
correlation matrix through:
\begin{equation}
\left\langle \delta A( z,\omega ) \left[ \delta A( z,\omega
^{\prime }) \right] ^{\dagger }\right\rangle =\frac LcS( z,\omega
) \delta ( \omega -\omega ^{\prime })\;. \label{def matriz S}
\end{equation}

Making use of  $\left\langle f( z,\omega ) f^{\dagger }( z^{\prime
},\omega ^{\prime }) \right\rangle =\frac LN2D\delta (
z-z^{\prime }) \delta ( \omega -\omega ^{\prime }) $,  where $
D$ is the atomic Langevin forces' diffusion matrix, and of Eq.
\ref{formal} one obtains:
\begin{eqnarray}
S( L,\omega)  &=&e^{BL}S( 0) ( e^{BL}) ^{\dagger }
+2\frac{N\eta^2}ce^{BL} \times\label{S propoagado} \\
&&\times\left[ \int_0^L e^{-Bz^{\prime
}}GDG^{\dagger } e^{-B^{\dagger }z^{\prime }} dz^{\prime }\right]
e^{B^{\dagger }L} \;, \nonumber
\end{eqnarray}
where we have omitted the dependence on $\omega $ for shortness
and taken $z=L$. 

The term proportional to the identity in $B$ commutes with all
other operators, so we can write:
\begin{eqnarray}
S( L,\omega)  &=&e^{KL}S( 0) e^{K^{\dagger }L}
+\frac{N\eta^2}ce^{KL}\times\label{S propagado  con K} \\
&&\times\left[ \int_0^L e^{-Kz^{\prime
}}Je^{-K^{\dagger }z^{\prime }}dz^{\prime }\right] e^{K^{\dagger
}L}\;, \nonumber
\end{eqnarray}
where $J=2GDG^{\dagger }$ and $K=\frac{N\eta^2}cGV$.

Let $X^{\prime }$ be a matrix satisfying:
\begin{equation}
-( KX^{\prime }+X^{\prime }K^{\dagger }) =J \;.\label{X prima}
\end{equation}
Then the integral in Eq. \ref{S propagado  con K} can be evaluated
and we get:
\begin{eqnarray}
S( L)  &=&e^{KL}S( 0) e^{K^{\dagger }L}
\label{S con X prima} \\
&&+\frac{N\eta^2}c( X^{\prime }-e^{KL}X^{\prime }e^{K^{\dagger
}L}) \;. \nonumber
\end{eqnarray}

Introducing $X=\frac{\Gamma X^{\prime }}L$ we have:
\begin{eqnarray} \label{S con C}
S( L)  &=&e^{C\Gamma GV}S( 0) e^{( C\Gamma GV) ^{\dagger
}}  \\
&&+C\left[ X-e^{C\Gamma GV}Xe^{( C\Gamma GV) ^{\dagger }}\right]\;, \nonumber
\end{eqnarray}
where $C=\frac{N\eta^2L}{c\Gamma }$ is the cooperativity parameter
\cite {DANTAN04,DANTAN05}.\\

Eq. \ref{S con C} gives the spectral density matrix $S( L,\omega )
$ after the atomic medium, given the incident field spectral
density $S( 0,\omega ) $. The first term on the right-hand side of
Eq. \ref{S con C} represents the semiclassical effect on light
fluctuations owing to the mean value of the atomic polarization in
response to the mean incident field. Such a term may lead to noise
reduction and cross polarization effects. The second term on the
right-hand side of Eq. \ref{S con C} represents the light noise
introduced by the atomic quantum fluctuations via the matrix $X$,
which is determined by the Langevin forces' diffusion matrix $D$
(Eq. \ref{X prima}). It always corresponds to \emph{noise
increase} (detrimental to squeezing). The calculation of the
diffusion matrix $D$ can be made with the help of the generalized
Einstein theorem \cite {SARGENTBOOK74} and the corresponding
expressions are given in the Appendix.\\

In order to be able to calculate noise spectra for arbitrary field
quadratures, we introduce the rotated field operators array $\delta
A(\theta ) \equiv \left\{ a_1e^{-i\theta},a_1^{\dagger}%
e^{i\theta },a_2e^{-i\theta },a_2^{\dagger }e^{i\theta }
\right\}^{\top}$. It can be immediately calculated as  $\delta A(
\theta ) =T$ $\delta A$, where $T$ is the $4\times 4 $ matrix with
$e^{-i\theta },e^{i\theta },e^{-i\theta },e^{i\theta }$ along its
main diagonal. In a similar way, fluctuations along two arbitrary
orthogonal polarization unit vectors $\hat{e}_1^{\prime }$ and $\hat{e}%
_2^{\prime }$ can be calculated as $\delta A ^{\prime}( \theta )^ =R$ $%
\delta A( \theta )$, where $R$ is the polarization-basis-change
matrix. Thus, the spectral density matrix $S_\theta
^{\prime }( L,\omega ) $ for a given quadrature angle and
arbitrary choice of the polarization basis can be evaluated as:
\begin{equation}
S_\theta ^{\prime }( L,\omega ) =TRS( L,\omega ) R^{\dagger
}T^{\dagger }\;.  \label{S rotado}
\end{equation}
The noise spectrum $s_{j\theta }^{\prime }$ for the quadrature
angle $\theta $ of the field with polarization $\hat{e}_j^{\prime
}$ $(j=1,2)$, as can be measured by homodyne detection, can be
evaluated from the matrix elements $S_{\theta \mu \nu }^{\prime }$
of $S_\theta ^{\prime }( L,\omega ) .$ Dropping, for shortness,
the $L$ and $\omega $ dependence  we have:
\begin{subequations}\label{terminos de ruido}
\begin{eqnarray}
s_{1\theta }^{\prime } &=&S_{\theta 11}^{\prime }+S_{\theta
12}^{\prime }+S_{\theta 21}^{\prime }+S_{\theta 22}^{\prime }\\
s_{2\theta }^{\prime } &=&S_{\theta 33}^{\prime }+S_{\theta
34}^{\prime }+S_{\theta 43}^{\prime }+S_{\theta 44}^{\prime }\;.
\end{eqnarray}
\end{subequations}
Other matrix elements of $S_\theta ^{\prime }$ not appearing in
Eqs. \ref{terminos de ruido} are related to the field correlations
between the two polarizations. 

\subsection{\label{Doppler}Velocity distribution}

So far we have considered an homogeneous sample of atoms at rest.
We will generalize now our calculation to a sample of moving atoms
with velocity $v_{z}$ in the direction of the light propagation
axis. The velocity distribution is $\mathcal{W}\left( v_z\right) $
obeying $ \int_{-\infty }^{+\infty }\mathcal{W}\left( v_z\right)
dv_z=1$. Then Eq. \ref{propagg linearizada}
becomes:
\begin{equation}
\label{Maxwell con velocidad}
\frac{\partial \left[ \delta A\left( z,\omega \right) \right] }{\partial z}%
=B^{\prime }\delta A+\frac{N\eta}c\int_{-\infty }^{+\infty
}\mathcal{W}\left( v_z\right) G\left( v_z\right) f_{v_z}dv_z \;,
\end{equation}
with
\begin{eqnarray}
B^{\prime } &=& i\frac \omega
c\Bbb{I}_{4}  \\
&&+ \frac{N\eta^2}c\int_{-\infty }^{+\infty }\mathcal{W}\left(
v_z\right) G\left( v_z\right) V\left( v_z\right) dv_z \nonumber \\
G\left( v_z\right)  &=&WM\left( v_z\right) ^{-1} \\
M\left( v_z\right)  &\equiv &-i\omega \Bbb{I}_{n}+\mathcal{A}(v_z) \;.
\end{eqnarray}

$\mathcal{A}(v_z)$ depends on $v_z$ through the velocity dependent
detuning $\Delta = \omega_{0}-\omega_{L}-kv_{z}$
(see Eq. \ref{HL con operador sigma}) and $V\left( v_z\right) $ depends on $v_z$ through the mean values $%
\left\langle x\left( v_z\right) \right\rangle $.\\

After formal integration of Eq. \ref{Maxwell con velocidad} one
has:
\begin{eqnarray}
\label{formal con velocidad}
\delta A\left( z,\omega \right) &=&
e^{B^{\prime }z}\left[ \delta A\left( 0,\omega \right)
+ \frac{N\eta}{c}\int_0^z e^{-B^{\prime}z^{\prime}}  \right. \\
& \times & \left. \int_{-\infty }^{+\infty }\mathcal{W} (v_z)
G(v_z) f( v_z,z^{\prime },\omega) dv_z dz^{\prime } \right]\;.
\nonumber
\end{eqnarray}

Since the Langevin forces for atoms with different velocities are
uncorrelated, we impose:

\begin{eqnarray}
\label{f langevin correlaciones con velocidad}
&&\left\langle
f\left( z,v_z,\omega \right) f^{\dagger }\left( z^{\prime
},v_z^{\prime },\omega ^{\prime }\right) \right\rangle \\
&& = \frac L{N\mathcal{W} \left( v_z\right) }2D \delta \left(
z-z^{\prime }\right) \delta \left( \omega -\omega ^{\prime
}\right) \delta \left( v_z-v_z^{\prime }\right) \;.\nonumber
\end{eqnarray}

Using Eqs. \ref{formal con velocidad} and \ref{f langevin
correlaciones con velocidad} one can obtain the spectral density
matrix for the output field as:

\begin{eqnarray}
\label{S Doppler} S(L)  &=&e^{K^{\prime }L}S( 0) e^{K^{^{\prime
}\dagger }L} \\
&&+\frac{N\eta^2}ce^{K^{\prime }L}\left[ \int_0^L e^{-K^{\prime
}z^{\prime }} J^{\prime } e^{-K^{\prime \dagger }z^{\prime
}}dz^{\prime }\right] e^{K^{\prime \dagger }L} \;, \nonumber
\end{eqnarray}

with
\begin{equation} \label{Kprima}
K^{\prime }L=C\Gamma \int_{-\infty }^{+\infty }\mathcal{W}\left(
v_z\right) G\left( v_z\right) V\left( v_z\right) dv_z \;,
\end{equation}
and
\begin{equation} \label{Oprima}
J^{\prime }=\int_{-\infty }^{+\infty }\mathcal{W}\left(
v_z^{\prime }\right) G\left( v_z^{\prime }\right) DG\left(
v_z^{\prime }\right) ^{\dagger }dv_z^{\prime }\;.
\end{equation}

The expression for the spectral density matrix given in Eq. \ref{S
Doppler} represents a generalization of Eq. \ref{S propagado con
K} that can be evaluated by the same method. The integrals in Eqs.
\ref{Kprima} and \ref{Oprima} can be evaluated numerically, given the
velocity distribution $\mathcal{W}( v_z)$.

\section{\label{resultados}Calculated noise spectra}

The model developed in Section \ref{modelo} allows us to calculate
the spectral density of the fields transmitted through an
homogeneous atomic sample of thickness $L$ and total number of
atoms $N$ under a wide range of conditions, i.e. arbitrary choices
of the atomic levels' angular momenta, driving field polarization,
and incident field mean intensity. In addition, the model allows
the arbitrary choice of the branching ratio $b$ and the
longitudinal magnetic field $\mathcal{B}$. However, in this paper
we will restrict ourselves to consider a reduced number of
parameters. No magnetic field is considered and the branching
ratio $b$ is taken as one (no transition to states external to the
two-level system). Nevertheless, one should notice that some
choices of $F_{g}$ and $F_{e}$ effectively describe open systems
since there are trapping (dark) states not coupled to the applied
field. The driving field polarization ($\hat{e}_1$) is chosen to
be linear along axis $x$, which means that vacuum enters the
system with $y$ polarization ($\hat{e}_2$). The $x$ polarized
driving field is assumed to be in a coherent state
$|\alpha\rangle$. The corresponding reduced Rabi frequency
(independent of polarization and atomic dipole orientation) is
$\Omega_r=2 \alpha \eta$,
taken real. Under these conditions, the incident (white noise)
spectral density matrix is \cite{VERNAC02}:

\begin{equation}
\label{S0} S( 0,\omega ) =\left(
\begin{array}{llll}
1 & 0 & 0 & 0 \\
0 & 0 & 0 & 0 \\
0 & 0 & 1 & 0 \\
0 & 0 & 0 & 0
\end{array}
\right) \;,
\end{equation}
in units for which the standard quantum noise limit corresponds to
one.

We have analyzed the output field fluctuations for the same set of
orthogonal polarizations ($x$ and $y$) used for the incident field
($R=\Bbb{I}_4$ in Eq. \ref{S rotado}).  Noise spectra for a
different choice of the polarization basis can be easily obtained
using the appropriate matrix $R$. 
We have calculated the spectra
for both the amplitude quadrature ($\theta =0$) and the phase
quadrature ($\theta = \pi /2$). In all calculations, the
relaxation rate $\gamma$ was taken as $\gamma = 0.01\Gamma$, a
realistic figure for experiments using alkali atoms' D lines. In
the following, we analyze different level schemes depending on the
angular momentum of the ground and excited states. We begin by
considering the case of an homogenous sample of atoms at rest (cold
atom sample). The effect of atomic motion is discussed at the end
of this section.

\subsection{\label{tla}$F_{g}=0 \rightarrow F_{e}=1$}

Choosing the quantization axis in the direction ($x$) of the
linear polarization of the applied field, this case reduces to a
pure two-level system in which the field couples the Zeeman sublevels
$|F_{g}=0, m_{g}=0 \rangle$ and $|F_{e}=1, m_{e}=0 \rangle$. The
$y$ polarization, along which vacuum is entering the atomic
system, is totally uncoupled to the driven transition. As a
consequence, no change takes place for the $y$ polarization mode
and this field emerges from the sample in the vacuum state.

\begin{figure}
\includegraphics[width=8.6cm]{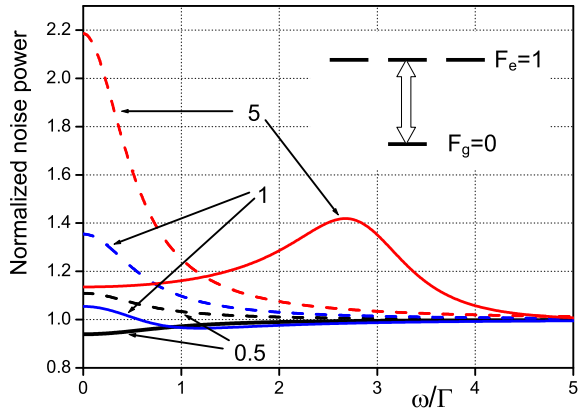}
\caption{\label{tladet0} (Color online) Transition $F_{g}=0
\rightarrow F_{e}=1$ (Inset: the level scheme is shown using $x$ as
quantization axis; hollow arrow: driving field). Noise spectra of
the amplitude (solid) and phase (dashed) quadratures of the
transmitted field with the same linear polarization as the
driving field, at zero detuning. The different values of the
reduced Rabi frequency $\Omega_r$ (in units of $\Gamma$) are
indicated ($C=1$, $\gamma =0.01 \Gamma$).}
\end{figure}

\begin{figure}
\includegraphics[width=8.6cm]{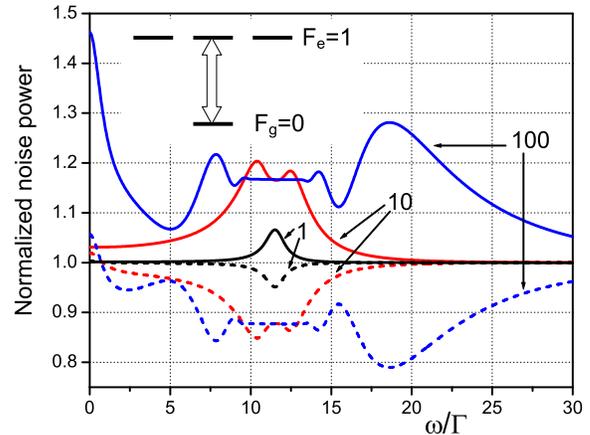}
\caption{\label{tladet10}  (Color online) Transition $F_{g}=0
\rightarrow F_{e}=1$. Noise spectra of the amplitude (solid) and
phase (dashed) quadratures of the field transmitted  with the same
linear polarization as the driving field for $\Delta = 10 \Gamma
= \Omega_r = 10 \Gamma$. The different values of the cooperativity
parameter $C$ are indicated ($\gamma =0.01 \Gamma$).}
\end{figure}

\begin{figure}
\includegraphics[width=8.6cm]{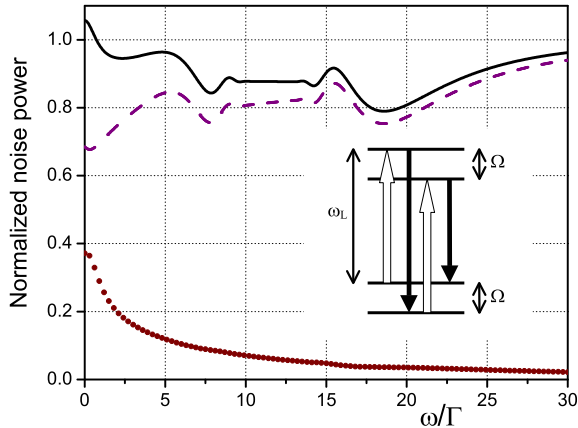}
\caption{\label{clasicoycuantico}  (Color online)  Transition
$F_{g}=0 \rightarrow F_{e}=1$. Solid: total noise spectrum for the
phase quadrature ($\Delta = \Omega_r = 10 \Gamma$, $\gamma =0.01
\Gamma$, $C = 100$). Dashed: semiclassical contribution (fist term
on the $r.h.s.$ of Eq. \ref{S con C}). Dotted:  contribution of
the atomic quantum fluctuations (second term on the $r.h.s.$ of
Eq. \ref{S con C}). Inset: Dressed-atom picture for the
four-wave-mixing process responsible for squeezing in two-level
atoms. Hollow arrows: laser mean field; solid arrows: noise
sidebands.}
\end{figure}

The results of our model are in agreement with previous work on
quantum fluctuations and squeezing for pure two-level atoms
\cite{COLLETT84,HEIDMANN85,HO87}. Fig. \ref{tladet0} presents the
noise spectra obtained for both quadratures in the case of
resonant atomic excitation ($\Delta =0$) for a rather thin sample
($C=1$), for different values of the reduced Rabi frequency
$\Omega_r$. Excess noise, with maximum value at zero frequency, is
obtained for the phase quadrature for all field intensities. The
amplitude quadrature presents squeezing centered at zero frequency
for small light field amplitude ($\Omega_r \lesssim 0.5\Gamma$).
As the Rabi frequency increases, the maximum squeezing shifts to
increasing nonzero frequencies and disappears as $\Omega_r >
\Gamma$. For large $\Omega_r$, the intensity fluctuations present
an excess noise peak centered at the {\em actual} Rabi frequency $\Omega_0$.
Notice that, since $\Omega_r$ is the reduced Rabi frequency, the
actual Rabi frequency associated to the specific two-level
transition is $\Omega_0 =(Q_{00,ge}^{0}) \Omega_r =
\frac{1}{\sqrt{3}} \Omega_r$. As discussed in \cite{HEIDMANN85},
amplitude squeezing  occurs in this case only in the limit of an
optically thin medium, where the amount of squeezing is linearly
dependent on $C$. However, at zero detuning, the mean field
absorption, which is not taken into account in our treatment, is
considerable and also linearly dependent on $C$. Consequently,
absorption of the incident field should prevent the observation of
squeezing in this regime. To avoid the effect of field absorption,
one should consider situations in which the generalized Rabi
frequency $\Omega=(\Delta^{2}+\Omega_0^{2})^{1/2}$ is large. Such
case corresponds to well resolved levels in the dressed-atom
picture \cite{COHENBOOK92}. Noise spectra calculated for $\Delta =
\Omega_r = 10 \Gamma$ are presented in Fig. \ref{tladet10}, for
different values of the cooperativity parameter. $C=1$ corresponds
to a thin atomic medium. The noise spectrum is essentially dominated
by a peak occurring at $\omega = \Omega$. Squeezing occurs for the
phase quadrature and, correspondingly, excess noise is present in
the amplitude quadrature. The observed squeezing is due to
four-wave mixing between the mean field at the carrier frequency
$\omega_{L}$ and the noise sidebands at frequencies $\omega_{L}
\pm \omega$ \cite{HO87}. In the dressed atom picture of the atoms
driven by the incident mean field, a double $\Lambda$ scheme
occurs, involving the absorption (emission) of two driving field
photons and the emission (absorption) of a photon from each of the
two sidebands (see inset in Fig. \ref{clasicoycuantico}). This
mechanism results in the buildup of correlations between the two
sideband fluctuations, leading to quadrature squeezing
\cite{BACHORBOOK98}. Such process is fully resonant with the
energy levels of the dressed-atom when $\omega =\Omega$.

As the medium's optical thickness increases, the noise feature
around $\omega = \Omega$ broadens considerably and presents
oscillations which are due to the frequency dependent phase
mismatch between the carrier mean field and the noise sidebands
\cite{HO87}. For large $C$ broadband squeezing is present.
More than $20\%$ squeezing is obtained for $C=100$.

We notice from Fig. \ref{tladet10} that, except for the largest
value of the cooperativity parameter $C$, there is no significant
variation of the noise power around zero frequency. This behavior
corresponds to closed two-level systems. Open two-level systems
show an increase of light fluctuations at low frequency.\\

The relative effect on the transmitted light fluctuations of the
semiclassical atomic response and that of the atomic quantum
fluctuations is illustrated in Fig. \ref{clasicoycuantico}. The
total spectrum for the phase quadrature noise and the
contributions corresponding to the two terms on the right-hand side of
Eq. \ref{S con C}, for $\Delta = 10 \Gamma = \Omega_r = 10 \Gamma$
and $C=100$, are shown. As expected, the squeezing is only due to
the semiclassical term contribution. For the chosen parameters,
the atomic quantum fluctuations introduce a
rather broad and smooth noise increase.\\

The fact that, for this level scheme, squeezing occurs for one
quadrature of the $x$ polarized field, while the vacuum
fluctuations incident along the $y$ polarization are unaffected,
implies entanglement between two orthogonally polarized modes
\cite{KOROLKOVA02}. This can be explicitly verified by noticing
that squeezing of the phase quadrature of field $1$ together with
vacuum fluctuations in field $2$ correspond to:

\begin{equation} \label{desigualdadsqueezing}
  \left[\Delta\left(\frac{a_{1}-a_{1}^{\dagger}}{i}\right)\right]^{2}
+\Delta(a_{2}+a_{2}^{\dagger})^{2}\leqslant 2\;.
\end{equation}

Let now consider $a_{+} = \frac{a_{1}+a_{2}}{\sqrt{2}}$ and $a_{-}
= \frac{a_{1}-a_{2}}{\sqrt{2}}$ the field operators corresponding
to the two orthogonal linear polarization at $45^{\circ}$ with
respect to the $x$ and $y$ axis. Then the operators
$X_{\mu}=\frac{a_{\mu}+a_{\mu}^{\dagger}}{\sqrt{2}}$ and
$Y_{\mu}=\frac{a_{\mu}-a_{\mu}^{\dagger}}{i\sqrt{2}}$ ($\mu=+,-$)
are two pairs of conjugate hermitian operators satisfying
$[X_{\mu},X_{\nu}]=0$, $[Y_{\mu},Y_{\nu}]=0$,
$[X_{\mu},Y_{\nu}]=i\delta_{\mu \nu}$. For these operators the
inequality in Eq. \ref{desigualdadsqueezing} becomes:

\begin{equation} \label{entantlement}
     \Delta(Y_{+}+Y_{-})^{2}+\Delta(X_{+}-X_{-})^{2}\leqslant 2\;,
\end{equation}
which is sufficient to demonstrate continuous variable
entanglement of the $+$ and $-$ polarization fields \cite{DUAN00}.

\subsection{\label{lambda}$F_{g}=1 \rightarrow F_{e}=0$}

\begin{figure}
\includegraphics[width=8.6cm]{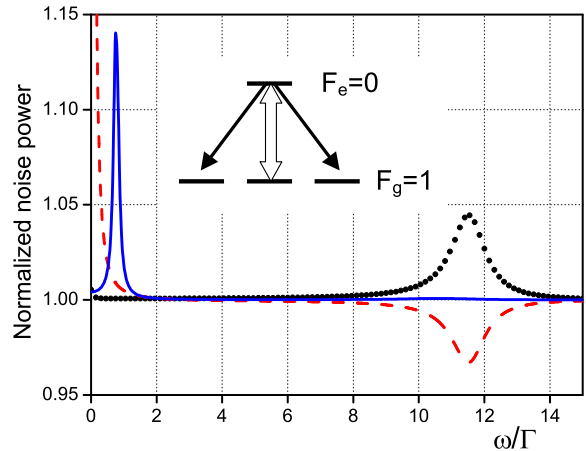}
\caption{\label{lambdadet10}  (Color online)  Transition $F_{g}=1
\rightarrow F_{e}=0$ (the level scheme is shown using $x$ as
quantization axis; hollow arrow: driving field; solid arrows:
spontaneous emission channels into the $y$ polarized mode). Noise
spectra of the amplitude (dotted) and phase (dashed) quadratures
of the transmitted field with the same linear polarization as
the driving field. The solid line corresponds to the noise of both
quadratures of the output field with polarization orthogonal to
the driving field ($\Delta = \Omega_r = 10 \Gamma$, $\gamma =0.01
\Gamma$, $C = 10$).}
\end{figure}

Excitation of a transition from a ground state with $F_{g}=1$ to a
$F_e=0$ excited state with a linearly ($x$) polarized field
corresponds to the coupling of the incident field to an open
two-level system. Choosing the quantization axis along $x$
determines that the incident field is coupled to the $|F_{g}=1,
m_{g}=0 \rangle$ to $|F_{e}=0, m_{e}=0 \rangle$ transition. The
excited state can decay into either of the ground state Zeeman
sublevels. Alternatively (in a different basis for the ground
state manifold), the configuration corresponds to the excitation
of one branch of a $\Lambda$ system while vacuum field is acting
on the second branch. A similar situation was studied in
\cite{DANTAN05}, in the case of a resonant excitation ($\Delta
=0$). Fig. \ref{lambdadet10} shows the noise spectra for both
quadratures of the output fields with $x$ and $y$ polarizations.
The spectra corresponding to the $x$ polarization are very similar
to those obtained for a closed two-level system (Fig.
\ref{tladet10}), except for the noise increase around zero
frequency. Such a low-frequency feature in the noise spectra
occurs in open transitions owing to the fluctuations introduced by
spontaneous decay out of the two-level system. In addition, an
increase in the noise power above the vacuum fluctuations is seen
for the $y$ polarized field. Both quadratures with this
polarization experience the same noise contribution due to
spontaneous emission from the excited state. As a consequence of
the light shift of the excited state produced by the driving
field, the noise for the $y$ polarization peaks at $\omega \neq
0$. In addition to this low frequency peak, a much smaller local
maximum occurs for $\omega \backsimeq \Omega$, almost invisible on
the scale of Fig. \ref{lambdadet10}.

\subsection{\label{X}$F_{g}=1/2 \rightarrow F_{e}=1/2$}

\begin{figure}
\includegraphics[width=8.6cm]{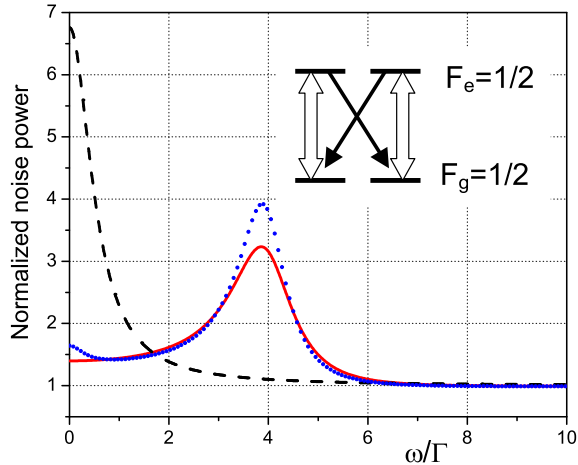}
\caption{\label{Xdet0}  (Color online) Transition $F_{g}=1/2
\rightarrow F_{e}=1/2$ (the level scheme is shown using $x$ as
quantization axis; hollow arrow: driving field; solid arrows:
spontaneous emission channels into the $y$ polarized mode).
Resonant noise spectra of the amplitude (solid) and phase (dashed)
quadratures of the transmitted field with the same linear
polarization as the driving field. Dotted: amplitude quadrature
noise spectrum of the polarization perpendicular to the driving
field, the spectrum of the phase quadrature for this polarization
coincides with the dashed line ($\Delta = 0$, $\Omega_r = 10
\Gamma$, $\gamma =0.01 \Gamma$, $C = 10$).}
\end{figure}

This configuration was analyzed in \cite{JOSSE03BIS} as a model
system for the study of squeezing via PSR. Fig. \ref{Xdet0}
presents the corresponding spectra for resonant excitation ($\Delta
=0$). The noise spectra for the driving field polarization are
analogous to those in Fig. \ref{tladet0} for a two-level
transition. In fact, for a choice of the quantization axis along
$x$, the present situation corresponds to two two-level
transitions coupled through spontaneous emission. Spontaneous
emission is also responsible for the injection of field
fluctuations for the $y$ polarization. Owing to the large symmetry
of this configuration, the amplitude noise spectrum for the $y$
polarization is equal to the phase noise spectrum of the $x$
polarization. The phase noise spectrum for the $y$ polarization
presents enhanced noise if compared to the amplitude quadrature of
the $x$ polarization.

\begin{figure}
\includegraphics[width=8.6cm]{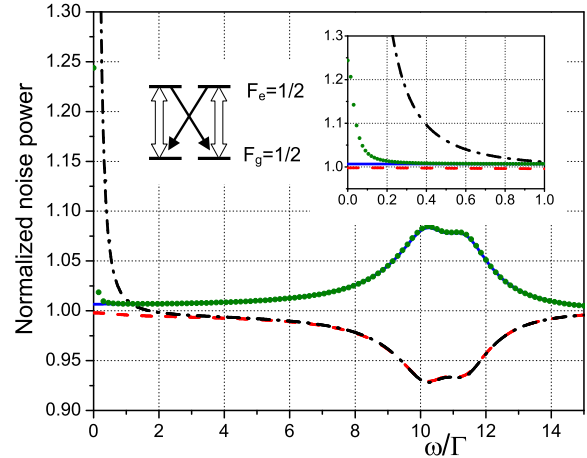}
\caption{\label{Xdet10}  (Color online) Transition $F_{g}=1/2
\rightarrow F_{e}=1/2$, $\Delta = 10\Gamma$. Amplitude (solid)
and phase (dashed) quadratures of the transmitted field with
the same linear polarization as the driving field. Amplitude
(dash-dot) and phase (dotted) quadratures
of the transmitted field with polarization perpendicular to the
driving field. Inset: expanded low-frequency range ($ \Omega_r =
10 \Gamma$, $\gamma =0.01 \Gamma$, $C = 10$).}
\end{figure}

The spectra for a non-resonant excitation ($\Delta = 10 \Gamma$)
are presented in Fig. \ref{Xdet10}. For $\omega \gtrsim \Gamma$
the spectra reproduce the features observed for non-resonant
excitation of two-level systems (see Fig. \ref{tladet10}). Nearly
identical noise spectra are obtained for the $x$ polarization
amplitude (phase) noise and the $y$ polarization phase (amplitude)
noise. This approximate symmetry is broken for low noise
frequencies (see inset on Fig. \ref{Xdet10}). Squeezing due to PSR
occurs in this system for nonzero detuning. In addition, the
inequality \ref{desigualdadsqueezing} is verified for $\omega \sim
\Omega$, giving rise to entanglement.

\subsection{\label{W}$F_{g}=1 \rightarrow F_{e}=2$}

We now consider a transition of the type $F_{g} >0 \rightarrow
F_{e}=F_{g}+1$. For this class of transitions, no dark state exists
within the ground level. At two-photon Raman resonance between
ground state Zeeman sublevels, coherence resonances occur that
correspond to enhanced absorption, that is electromagnetically
induced absorption (EIA) \cite{AKULSHIN98,LEZAMA99}. The
absorption spectra of a transition of this type driven by a strong
classical field were examined in  \cite{LIPSICH00}. If
a driving field of circular polarization is used, owing to optical
pumping, the system approximates a pure two-level system. However,
if linear polarization is used for the driving field, while the
system is probed along the orthogonal polarization, the absorption
spectrum presents a rather complex structure as a result of the
different light shifts experienced by the Zeeman sublevels. A
simple picture of this effect in terms of the dressed atom model
is presented in \cite{LIPSICH00}.

\begin{figure}
\includegraphics[width=8.6cm]{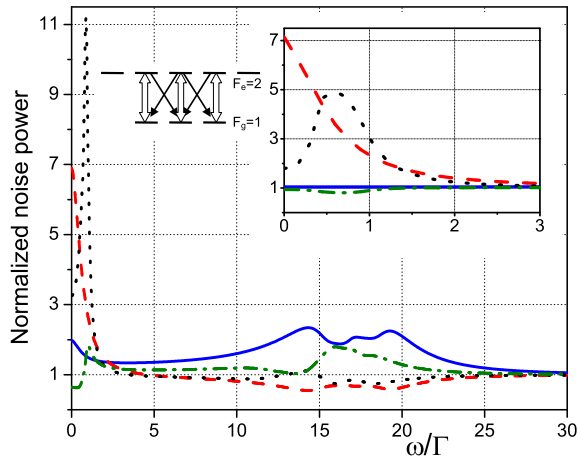}
\caption{\label{Wdet40}  (Color online) Transition $F_{g}=1
\rightarrow F_{e}=2$ (the level scheme is shown using $x$ as
quantization axis; hollow arrow: driving field; solid arrows:
spontaneous emission channels into the $y$ polarized mode). Non-
resonant noise spectra ($\Delta = 10\Gamma$). Amplitude (solid)
and phase (dashed) quadratures of the transmitted field with the
same linear polarization as the driving field. Amplitude
(dash-dot) and phase (dotted) quadratures of the transmitted field
with polarization perpendicular to the driving field. Inset:
expanded low-frequency range ($\Omega_r = 40 \Gamma$, $\gamma
=0.01 \Gamma$, $C = 100$).}
\end{figure}

The transmitted field noise spectra for the transition $F_{g}=1
\rightarrow F_{e}=2$ are presented in Fig. \ref{Wdet40} for
nonzero detuning ($\Delta = 10 \Gamma$) and large Rabi frequency
($\Omega_r = 40 \Gamma$). A thick optical medium is considered
($C=100$). For $\omega > \Gamma$, the spectra corresponding to the
driving field polarization are similar to the pure two-level
spectra shown in Fig. \ref{tladet10}. Unlike the pure two-level
system, noise is introduced by spontaneous emission into the $y$
polarized field. However, PSR squeezing is nevertheless present on
the amplitude quadrature for this polarization. In addition, the
inequality in Eq. \ref{desigualdadsqueezing} is verified resulting
in entanglement.

At low frequencies ($\omega \lesssim \Gamma$), the noise spectra
present features that are specific of such a multilevel system.
Excess noise peaking at $\omega =0$ occurs for the two quadratures
of the driving field polarization. The orthogonal polarization
presents features which are peaked at $\omega > 0$ owing to the
different Zeeman sublevel light shifts \cite{GRISON91,LIPSICH00}.
Notice the significant squeezing near zero frequency of the phase
quadrature.

\subsection{Effect of atomic motion}

\begin{figure}
\includegraphics[width=8.6cm]{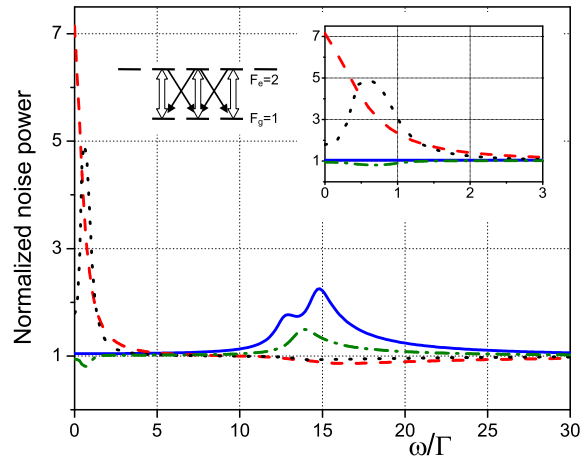}
\caption{\label{con Doppler}  (Color online) Transition $F_{g}=1
\rightarrow F_{e}=2$. Noise spectra for a Maxwell-Boltzmann atomic
velocity distribution and driving field tuned to resonance with
zero velocity atoms. Amplitude (solid) and phase (dashed)
quadratures of the transmitted field with the same linear
polarization as the driving field. Amplitude (dash-dot) and
phase (dotted) quadratures of the transmitted field with
polarization perpendicular to the driving field. Inset: expanded
low-frequency range ($\Omega_r = 40 \Gamma$, $\gamma =0.01
\Gamma$, $C = 100, \Gamma_{Dopp} =90\Gamma$).}
\end{figure}

We will now illustrate the effect of atomic motion on the noise
spectra for an atomic sample with a Maxwell-Boltzmann velocity
distribution. Having in mind the case of rubidium vapor at room
temperature, we have used a Doppler width (FWHM) $\Gamma_{Dopp}
=90\Gamma$. We consider as an example the $F_{g}=1 \rightarrow
F_{e}=2$ transition (same as in Fig. \ref{Wdet40}).

Figure \ref{con Doppler} shows the noise spectra for both field
quadratures in the case of a driving field resonant with the zero
velocity atoms and a Rabi frequency $\Omega_r = 40
\Gamma$ for $C=100$. Notice that, although the average detuning is
zero in this case, the noise spectra are reminiscent of those
obtained for atoms at rest with nonzero detuning. A local
maximum (minimum)is observed for the amplitude (phase) quadrature
noise of the driving field polarization for $\omega
\simeq\Omega_{0}$. The squeezing of the phase quadrature
remains significant ($\sim 15\%$) in spite of the spreading of
the field detuning caused by the Doppler effect. Notice that the
squeezing of the amplitude quadrature noise of the $y$ polarized
field is not completely suppressed by the velocity distribution
nor are the low frequency spectral features for both
polarizations.

\section{\label{conclusiones}Conclusions}

We have developed a theoretical model supporting the numerical
calculation of the noise spectra of light traversing an
optically thick atomic medium, under a wide range of conditions.
The model takes into account the complete Zeeman sublevel
structure of the atomic transition and consequently the arbitrary
polarization of the light field. In this paper, we have restricted
our analysis to the case of a linearly polarized driving field
while considering that the Zeeman sublevels are degenerate in the
absence of light (zero magnetic field). We have calculated the
noise spectra  for the amplitude and phase quadratures of the
transmitted light with the same and the orthogonal polarization
relative to the driving field.

For atoms at rest, as can be produced in magneto-optical traps,
the dominant feature of the transmitted field fluctuation spectra
can be traced back to pure two-level atom effects
\cite{HEIDMANN85}. For nonzero optical detuning, excess noise is
introduced in the amplitude quadrature while squeezing occurs for
the phase quadrature. The noise variations are maximum for $\omega
=\Omega$, as a consequence of resonant four-wave mixing of the
mean field and noise sidebands with the dressed-atom level
structure. The propagation through the optically thick medium
results in broadening and oscillatory structure of the noise
spectrum owing to phase mismatch between the carrier and
the fluctuation sidebands. Up to $30\%$ squeezing occur for the
parameter values used in the simulations, chosen in order to
correspond to typical experimental conditions, involving CW lasers
and alkali atoms. The figure of $C=100$ used in several of the
present calculations corresponds to a realistic figure for
magneto-optically trapped atom clouds.

In addition to pure two-level atom features, novel effects
are present as a consequence of the Zeeman sublevel structure. Except when
$F_{g}=0$, spontaneous emission introduces noise into the
polarization orthogonal to the driving field. Notwithstanding the
random field fluctuations caused by spontaneous emission, nonlinear
interaction between noise sidebands on both polarizations, may
result (as in Figs. \ref{Xdet10} and \ref{Wdet40}) in vacuum
squeezing for the orthogonal polarization. For the conditions used
in our calculations, the noise introduced by the atomic quantum
fluctuations does not completely mask the orthogonal polarization
vacuum squeezing. Furthermore, for transitions involving multiple
Zeeman sublevels, new structures appear in the noise spectra at
low frequency. They come from the different Stark
shifts experienced by the Zeeman sublevels in the presence of the
driving field. Such low frequency structures are reported here for
the first time.

The study presented here constitutes a necessary step towards a
satisfactory understanding and control of atom-light interaction
at the quantum level. This control can be applied to matter-light
interfaces for quantum information purposes, as the entanglement
predicted here implies. Further developments may include the
influence of nearby transitions as well as that of the spatial
structure of the light mode and atomic sample. Pulse propagation
effects should also be considered. Such theoretical approach needs
to be complemented with experimental tests, in particular using cold
atomic samples. Work in this direction is currently underway.

\appendix
\section{Calculation of the diffusion coefficients matrix}

The Heisenberg-Langevin equation (Eq. \ref{HL con operador sigma})
can be written in the form:

\begin{subequations}
\label{HL con drift}
\begin{eqnarray}
\frac{d\sigma _{ij}}{dt} &=&\mathcal{D}\left( \sigma _{ij}\right)
+f_{ij} \\
\mathcal{D}\left( \sigma _{ij}\right)  &\equiv &\sum_{kl}\mathcal{A}%
_{ij,kl}\sigma _{kl}+\gamma \sigma _{ij}^0 \;,
\end{eqnarray}
\end{subequations}
where $\mathcal{A}\equiv \left\{\mathcal{A}_{ij,kl}\right\} $ is the
matrix appearing in Eq. \ref{linearizada deltax}. The Langevin forces obey: $%
\left\langle f_{ij}\left( z,t\right) f_{kl}^{\dagger }\left(
z^{\prime },t^{\prime }\right) \right\rangle =\frac
LN2D_{ij,kl}\delta \left( z^{\prime }-z^{\prime \prime }\right)
\delta \left( t-t^{\prime }\right) $. The diffusion matrix
$D=\left\{ D_{ij,kl}\right\} $ can be calculated with the help of
the generalized Einstein relation
\cite{SARGENTBOOK74,COHENBOOK92}:

\begin{equation}
2D_{ij,kl}=\left\langle \mathcal{D}\left( \sigma _{ij}\sigma
_{kl}^{\dagger }\right) -\mathcal{D}\left( \sigma _{ij}\right)
\sigma _{kl}^{\dagger }-\sigma _{ij}\mathcal{D}\left( \sigma
_{kl}^{\dagger }\right) \right\rangle\;. \label{E generalizada}
\end{equation}

From Eq. \ref{HL con drift}b, by making use of $\sigma
_{ij}^{\dagger }\sigma _{kl}\equiv \left| i\right\rangle
\left\langle j\right| $ $\left| l\right\rangle \left\langle
k\right| =\sigma _{ik}^{\dagger }\delta _{jl}$ and
$\mathcal{A}_{ij,mn}=\mathcal{A}_{ji,nm}^{*}$ we get:

\begin{eqnarray}
\label{Dijkl}
\left\langle \mathcal{D}\left( \sigma _{ij}\sigma _{kl}^{\dagger
}\right)
\right\rangle  &=&0 \\
\left\langle \mathcal{D}\left( \sigma _{ij}\right) \sigma
_{kl}^{\dagger }\right\rangle
&=&\sum_m\mathcal{A}_{ij,km}\left\langle \sigma _{lm}\right\rangle
+\gamma \sigma _{ij}^0\left\langle \sigma
_{lk}\right\rangle   \nonumber \\
\left\langle \sigma _{ij}\mathcal{D}\left( \sigma _{kl}^{\dagger
}\right) \right\rangle  &=&\sum_m\mathcal{A}_{lk,mi}\left\langle
\sigma _{mj}\right\rangle +\gamma \left\langle \sigma
_{ij}\right\rangle \sigma _{lk}^0 \;. \nonumber
\end{eqnarray}
The first of Eqs. \ref{Dijkl} results from the stationarity of the
system. The mean values $\left\langle \sigma _{ij}\right\rangle $
are obtained from the steady state solution of Eq. \ref{HL con
drift}a.


\end{document}